# New Aspects of Temporal Dispersion in High Resolution Fourier Pulse Shaping: A Quantitative Description with Virtually Imaged Phased Array Pulse Shapers

**V. R. Supradeepa,[*] Ehsan Hamidi,  Daniel E. Leaird, and Andrew M. Weiner**

*School of Electrical and Computer Engineering, Purdue University, 465 Northwestern Avenue,*

*West Lafayette, Indiana 47907, USA*

*[*]Corresponding author: supradeepa@purdue.edu*

We report new aspects of temporal dispersion in Fourier pulse shapers which contain spectral dispersers with a nonlinear frequency to space mapping. These effects are particularly important in high resolution operation since high resolution dispersers typically exhibit pronounced nonlinear angular dispersion over relatively small bandwidths.  In this paper we present a general discussion of the new effects followed by quantitative analysis and experimental verification for pulse shapers which utilize a virtually imaged phased array (VIPA) as the spectral disperser. Compared to the well known 4f configuration, our results demonstrate a substantial modification to the placement of the optical components required to obtain zero temporal dispersion. Furthermore, spectral phase variations associated with nonzero dispersion coupled with contributions from multiple diffraction orders are shown to give rise to a dramatic new spectral interference effect, which can be used to monitor temporal dispersion purely in the spectral domain. We expect the effects we present in this paper to become prominent even for more conventional diffraction grating based pulse shapers for bandwidths sufficiently large that nonlinear spectral mapping becomes strong.

## 1. Introduction

Femtosecond pulse shaping is a well established technique which allows for synthesis of complex user defined optical waveforms by parallel manipulation of the complex optical spectrum [1]. In its conventional implementation, a spectral disperser (e.g. diffraction grating) is used which spatially disperses different frequency components which are then focused on to a spatially encoded mask with user defined amplitude and phase information. This can be a programmable spatial light modulator or a fixed mask made using lithographic techniques. This information gets encoded onto the complex spectrum of the light source and on recombination leads to synthesis of user defined waveforms. An important attribute desirable from the pulse shaping apparatus is that there should be no additional amplitude or phase effects other than the ones which are user defined. Another way of looking at this is, in the absence of a mask, the output and input fields should be identical. Usually with reasonable choice of components, the spectral amplitude effects due to the apparatus are minimal; however it is more involved when it comes to spectral phase effects. The origins of phase effects lie in the well known fact that whenever there is angular dispersion of spectral components there is a possibility of temporal dispersion, which in the frequency domain corresponds to additional spectral phase of second order and higher [2, 3]. In pulse shapers since there is spectral dispersion and recombination before and after masking respectively, it becomes necessary to ensure that the apparatus does not introduce any additional temporal dispersion of the input light. In such a configuration the input and output in the absence of masking will be identical and this is called the zero dispersion configuration.

The most common spectral disperser used in pulse shaping is a diffraction grating, for which the zero dispersion configuration is well characterized and known as the 4-F configuration [1]. In this configuration the lens is located a one focal length (F) distance behind the diffraction grating and a one focal length distance in front of the Fourier plane where the mask is placed. In either of the two main schemes which exist for recombination [either (1) reflective geometry, where a mirror at the Fourier plane reflects the light back in the same path as the incident light, or (2) transmission geometry, where a $2^{nd}$ identical lens and grating pair is used after the Fourier plane], light travels a distance of 4F between its first and second incidence on a grating. The 4-F configuration provides a dispersion-free configuration when using diffraction gratings over broad bandwidths and has been used extensively.

Recently there has been significant research in pulse shaping utilizing other spectral dispersers with higher spectral resolution than diffraction gratings such as Arrayed Waveguide Gratings (AWGs) [4] and Virtually Imaged Phased Arrays (VIPAs) [5]. Note that there is an inverse relation between spectral domain and temporal domain in pulse shaping: a broad bandwidth corresponds to fine temporal features while high spectral resolution corresponds to wide temporal windows. Hence a high spectral resolution is advantageous in applications like optical communications and RF photonics, where the temporal apertures involved are wide (for e.g. ns time scales). For other applications like spectroscopy, a high spectral resolution is inherently desirable. One of the main candidates as a high resolution disperser has been the virtually imaged phased array (VIPA) [5] which is a side entrance Fabry-Perot etalon device which achieves high resolution spectral dispersion by multiple beam interference (a schematic is shown in fig 1 and will be discussed in more detail later). VIPAs have already been adopted for various applications, including dispersion and polarization mode dispersion compensation in lightwave systems [6-8], radio-frequency photonic waveform generation and filtering [9-10], and as components in high resolution spectrometers for spectroscopy [11-12]. These components have also been applied for the development of hybrid, two dimensional pulse shapers, which combine the benefits of both VIPAs and gratings - VIPAs for high resolution and gratings for broad bandwidth - to achieve very high complexity waveform synthesis and filtering operations [13-14]. A key advantage VIPAs have over other dispersers is the achievable spectral resolutions. VIPAs have been shown to achieve sub-GHz resolutions while other high resolution dispersers like the AWG are still an order higher than that. This enhanced resolution is what makes VIPAs particularly appleaing. Another advantage VIPAs have over AWGs for femtosecond pulse shaping in particular is that, by the nature of AWGs originally being intended as spectral demultiplexers for WDM systems, it introduces spectral structure (reduced transmission at the edges of each channel). This is unimportant for communication applications, where each channel is independent, but for femtosecond pulses, which have very wide bandwidths, each channel is related. The introduction of spectral structure causes additional time domain effects which can be avoided with VIPA based systems. An advantage AWG based systems might have over VIPA shapers is that owing to their integrated nature they are more compact.

In order to use a VIPA or other high resolution disperser in a pulse shaping apparatus, it becomes necessary to understand the temporal dispersion characteristics and look for the zero dispersion configuration. As we show here, for a VIPA-based pulse shaper, the 4-F configuration does not have zero dispersion; significant modification to the placement of the lens is necessary to obtain zero dispersion. A main factor contributing to this modified temporal dispersion behavior is the pronounced nonlinearity in the frequency to space mapping by a VIPA spectral disperser. This is in contrast to gratings which have much more linear mapping over substantial bandwidths. In presence of this nonlinear mapping, additional effects need to be taken into consideration in analyzing temporal dispersion. In this work we will describe these new temporal dispersion effects and then develop a quantitative model for the temporal dispersion in a VIPA based pulse shaper. We expect the effects we present in this paper to become prominent even for more conventional diffraction grating based pulse shapers for bandwidths sufficiently large that nonlinear spectral mapping becomes strong.

New effects described in this paper for VIPA-based pulse shapers include the following:

- The pulse shaper exhibits strong dispersion in the 4F configuration.

- The periodicity of this dispersion converts a single input pulse into a long sequence of pulses separated in time by the inverse of the VIPA free spectral range.

- In the frequency domain, this dispersion - coupled with contributions from multiple diffraction orders - leads to the appearance of interference fringes in the power spectrum.

- Zero dispersion is obtained only with substantial change of the separation between spectral disperser and lens.

- Furthermore, our results furnish insight that allows for simultaneous minimization of dispersion and insertion loss.

This paper is organized into the following sections. In section 2 we will describe some observations made with a VIPA based pulse shaper that motivate the need for further understanding of dispersion. In section 3, we will give a qualitative explanation of the new temporal dispersion aspects for a generic spectral disperser with a nonlinear frequency to space mapping. In sections 4 and 5, we respectively provide a rigorous quantitative description for VIPA based pulse shapers and describe experiments that confirm our analysis. In section 5 we also discuss how modulation introduced into the optical spectrum through interference of multiple diffraction orders may be used as a simple means to sense and accurately monitor temporal dispersion. In section 6 we will comment on the close connection between the dispersion and loss behavior of VIPA shapers and conclude.

## 2. Some observations

Fig 2 shows an experiment where a VIPA based pulse shaper is aligned in a 4-F configuration without the application of any mask. This is to get an idea of how much dispersion is present in the 4-F configuration, which is the dispersion free configuration for a grating based pulse shaper. Fig. 2(A) shows the input pulse with a pulse width of ~100fs to the pulse shaper and fig. 2(B) shows a section of the input spectrum corresponding to one VIPA free spectral range (FSR) , which in this case is 200GHz (~1.6nm at 1550nm). The input pulse is a bandwidth limited pulse and the spectrum is flat. Since the VIPA is a Fabry Perot etalon based device, the frequency response is periodic at the free spectral range; in experiments for which the input bandwidth is greater than the FSR, the modifications introduced by the VIPA are periodic at the FSR. Consequently, plotting one FSR of spectral data is sufficient. At the output (fig 2(C)) we see that the intensity cross-correlation corresponds to a pulse train under a wide temporal envelope, significantly different compared to the input. This results from the large temporal dispersion introduced by the pulse shaper. The reason why we observe a pulse burst instead of a single broadened pulse is connected to the use of a short input pulse with bandwidth much wider than the FSR of the spectral disperser. So any modification to the complex spectrum repeats at periods of one FSR. If the pulse shaper introduces temporal dispersion, which in the frequency domain is a spectral phase variation, this phase variation is periodically repeated. The net effect is a pulse train where the individual pulse widths remain similar to the input pulse duration width, but where pulses repeat under a temporal envelope whose width is dependent on the amount of dispersion introduced. We see from the width of the temporal envelope that at the 4-F configuration a ~100fs input pulse is broadened to more than 100ps; which is very far indeed from being dispersion free. Another interesting observation is that the spectrum of Fig. 2(D)) now exhibits deep ripples. This ripple structure repeats periodically at the VIPA FSR. Since temporal dispersion is purely a spectral phase effect, such spectral amplitude effects require explanation. As mentioned earlier, this amplitude modulation arises due to interference between multiple diffraction orders of the VIPA in the presence of nonzero temporal dispersion.

Some additional observations help provide insight into new factors contributing to temporal dispersion. Fig 3(A) shows a schematic of a VIPA pulse shaper aligned in the 4-F configuration. Three important alignment parameters, indicated by the arrow marks, are varied: 1) the VIPA-lens distance (by a few cm), 2) the transverse position of the lens with respect to the VIPA, (by a few mm) and 3) the tilt angle of the mirror (by a few tenths of a degree). Figs. 3(B)-3(G) show the output time domain profiles, measured via intensity cross-correlation, and the corresponding output power spectra obtained when the transverse position of the lens was varied. Similar effects were observed with variation of VIPA-lens distance and mirror tilt angle. Some key points to note are:

- Varying the VIPA to Fourier lens distance is expected to change dispersion, similar to the case with gratings. However, the variation of the transverse position of the lens and the tilt of the return mirror do not affect the temporal dispersion for a grating based system. In the case of a VIPA based shaper, these alignment aspects contribute significantly to dispersion.

- The above observation contains the key ingredient to understand these new dispersion effects. As we will analyze later, varying the transverse position of the lens or the mirror tilt introduces a linear spatial phase (wavefront tilt) in the optical system. The time domain properties though are related to the spectral phase. In the case of a linear spectral dispersion device (like a grating for reasonable bandwidths), a linear spatial phase causes a linear spectral phase. However when there is pronounced nonlinear spectral dispersion, a linear spatial phase leads to higher order spectral phase and hence dispersion. New dispersion effects arise due to linear spatial phase (which are ignored for grating systems) coupled with nonlinear spectral dispersion.

- The temporal envelope width (which is a measure of dispersion) is observed to vary inversely with the average period of the spectral ripple. The spectral ripple disappears when zero dispersion is achieved. The close connection between temporal dispersion and the spectral ripple will become clear from our analysis.

## 3. Qualitative analysis

The corresponding parameter to temporal dispersion in the frequency domain is spectral phase of order two and higher [3]. A quantitative analysis of the amount of dispersion introduced involves evaluating the amount of spectral phase introduced by the pulse shaper which in turn depends on the nature of the spectral disperser used and the geometry of the optical configuration. However, before going into a detailed analysis, we introduce a simpler method to visualize the presence of temporal dispersion in pulse shapers. Using this method we will discuss how the origin of dispersion in VIPA or other high resolution pulse shapers differs from that in grating based shapers.

Fig 4(A) shows a schematic of a grating based pulse shaper (shown in a transmission geometry for easy visualization) aligned away from a zero dispersion configuration. Here, let us assume that the incident beam size is small. After spectral dispersion by the grating, the lenses first image different frequency components at different positions on the Fourier plane and then recombine them. However if the sum of the grating – lens distances differs from 2F (here shown for detuning of only the second $2^{nd}$ grating-lens pair), then before recombination different frequencies end up travelling different distances - which corresponds to temporal dispersion. This however will not be the case if the sum of the grating – lens distances is exactly equal to 2F (known as the 4-F configuration).

However in the case of a VIPA, it is a different situation. Fig 4(B) shows the schematic of spectral dispersion by the VIPA. An air spaced VIPA with a thickness $t$ is considered. An input Gaussian beam is coupled into the VIPA at an angle of $\theta_i$ through a nonreflecting window on the front side. Usually, the front side of the VIPA has ~100% reflectivity (except for the window) while the back side is partially reflecting (between 0.95-0.98). Multiple reflections occur inside the VIPA etalon; due to the partially reflecting nature of the back surface, at every reflection a fraction of the beam escapes as shown in the figure (see also Fig. 1). This phenomenon is best modeled by using virtual sources having Gaussian profiles to represent the successive beams, having longitudinal and transverse separations between successive sources of

$$\Delta z = 2t \cos \theta_i \qquad (1)$$

$$\Delta\xi = 2t\sin\theta_i \qquad\qquad (2)$$

These multiple virtual sources interfere to create the spectral dispersion. In this situation, unlike the previous case with the grating, spectral dispersion seems to originate from a line rather than a point. So it is clear that a 4-F configuration defined using the physical location of the first virtual source (which is the point where the incident beam first exits from the back surface of the VIPA) will not be the zero dispersion configuration. Fig 4(C) shows a schematic of a configuration of the VIPA shaper in which the first virtual source is not only located at a distance of F from the lens but is also moved below the optic axis - which as we will see later is necessary to achieve low loss. Now the reflected set of sources imaged twice through the lens is separated longitudinally from the input virtual sources. Hence similar to Fig. 4(A) there is now additional propagation distance, from which we expect further temporal dispersion to arise. So the configuration of Fig. 4(C) is not the zero dispersion configuration for the VIPA shaper. From this discussion we can guess that for the zero dispersion configuration, the line defined by the incident set of sources and the reflected set of sources should overlap exactly. This situation is shown in figure 4(D). As indicated in the figure, let the first virtual source be at a transverse distance of 'L' from the optic axis, and to generalize, let the mirror angle be taken as an additional parameter $\theta_m$. Also let the first virtual source be closer to the lens than F by distance d. To ensure overlap between incident and virtual sources, we obtain the condition

$$L + \frac{F}{2}\tan 2\theta_m - d\tan\theta_i = 0 \qquad\qquad (3)$$

We will prove rigorously in the next section that this indeed is the zero dispersion configuration for a VIPA based pulse shaper. At this point we can also comment on the zero dispersion configuration for a grating based shaper when using beams which are not small. In this case too, diffraction happens from a line, and we expect that the overlap of the incident beam (on the grating surface) and the reflected beam imaged twice by the lens would ensure zero dispersion. It is easy to see that this condition is satisfied when the focal point of the lens is on the diffracting surface of the grating.

Let us now discuss the spectral manifestation of these effects. By the nature of the imaging operation, if the grating –lens distance is not equal to the focal length of the lens, then on the Fourier plane a quadratic spatial phase is introduced [16] (i.e. in the Fresnel approximation, for incident light from a point source which is not at F from the lens, the wave front on the Fourier plane is not flat but has a quadratic shape). Since different frequency components occupy different spatial positions on the Fourier plane, a spatial phase will cause a spectral phase dependent on the frequency to space mapping. For the grating, over a reasonable bandwidth one may usually approximate the frequency to space mapping as linear, so a quadratic spatial phase becomes a quadratic spectral phase. This is the origin of the temporal dispersion in grating based shapers. If the spatial quadratic phase is made zero, which occurs in the 4-F configuration, the spectral quadratic phase and the temporal dispersion are also zero. The point to note here is that the spectral phase depends on two things – the spatial phase and the frequency to space mapping.

In case of gratings and other spectral dispersers which we approximate as linear, we don't usually consider sources of spatial linear phase, since that results only in linear spectral phase, which in the time domain translates only to a pulse delay (which is usually irrelevant). However in the case of high resolution spectral dispersers like the VIPA, there is pronounced nonlinearity in the frequency to space mapping (see Fig. 5 for an example showing clear quadratic frequency to space dependence for a VIPA (from [17])). In this situation even a linear spatial phase becomes relevant. Figures 6(A) and 6(B) show representative cartoons for the case of linear spectral dispersers and nonlinear spectral dispersers in the presence of linear spatial phase. The points represent equally spaced frequencies. In the case of linear spectral dispersion (Fig. 6(A)), the phase differences between equally spaced frequencies are constant, indicating absence of higher order phase terms. However, in case of nonlinear spectral dispersion (Fig. 6(B)), we see that the phase differences are not constant, indicating the presence of higher order spectral phase. Practically, linear spatial phase may arise due to a number of reasons. A common reason is beam asymmetry with respect to the Fourier lens, which may occur due to disperser specific reasons, alignment constraints or even misalignment. With regard to a VIPA, in the normal mode of operation a beam asymmetry is required in order to achieve low loss, which forces an unavoidable linear spatial phase. We will discuss this further in the next section.

To understand the origin of the ripples in the spectral domain, let us now consider simultaneous nonlinear spectral dispersion, linear spatial phase, and multiple diffraction orders, as sketched in Fig 7. The phase difference between the components of a given frequency in two different orders determines the amplitude of their coherent sum. This phase difference now becomes frequency dependent, leading to frequency dependent amplitude variations (spectral ripples). In the presence of a large temporal dispersion, the phase difference between adjacent orders varies more rapidly with respect to frequency, leading to more rapid spectral variations. This is in qualitative accord with the observed inverse relation between spectral ripple period and temporal dispersion. A point we would like to stress here is that the origin of pulse broadening in the time domain is the underlying spectral phase. In the presence of multiple diffraction orders, this also introduces ripples into the spectrum. If all the diffraction orders except one are blocked, the spectral ripples disappear but the dispersion in the time domain remains.

## 4. Quantitative analysis for VIPA based pulse shapers

In this section we will calculate the new dispersion effects discussed qualitatively above for VIPA based pulse shapers and prove rigorously the zero dispersion condition we suggested in section 3. The analysis in this section is built on the framework introduced in [17] to analyze the demultiplexing properties of the VIPA. Figure 8 shows a general configuration of a VIPA pulse shaper that we will use in the analysis. We use $\xi$ and $z$ to represent the transverse and longitudinal position coordinates in the object plane and $x$ to represent the transverse coordinate in the Fourier plane.

In reference [17] the distance between the first virtual source to the Fourier lens is taken as the focal length of the lens (F). As we will see later, for this case no quadratic spatial phase is observed at the Fourier plane. However as we discussed previously, having zero quadratic spatial phase is not a sufficient condition to have zero temporal dispersion since even a linear spatial phase can cause higher order spectral phases. Here we will allow the VIPA – lens distance (specifically, the distance between the lens and the first virtual source) to differ from the focal length by an amount d. For d>0, the distance between the first virtual source and the lens is less than the focal length. We also allow the first virtual source to be offset from the optic axis by $L$ and the folding mirror to be tilted by an angle $\theta_m$. The distance between the mirror and the Fourier lens is maintained at F since this ensures both minimum loss and best spectral resolution for pulse shaping.

Assuming that the power loss from the top end of the etalon is negligible, we can model the field on the Fourier plane as a superposition of contributions from an infinite number of virtual sources whose profiles go as

$$E_n(\xi) = (R)^n E_0 \exp(-\frac{(\xi - \Delta\xi_n + L)^2}{w_0^2}) \,(4)$$

where $R$ is the effective reflection coefficient per bounce inside the VIPA (around $0.95 - 0.98$), $\Delta\xi_n$ gives the shift in the transverse position of the nth virtual source given by

$$\Delta\xi_n = n*2t\sin\theta_i \qquad\qquad (5)$$

and $w_0$ is the focused beam radius at the back surface of the VIPA for the input beam. The field on the Fourier plane can be written as [17]

$$E_{out}(x,k) \propto \sum_{n=0}^{\infty} \exp(-ik(\Delta z_n - d))\exp(ik\frac{(\Delta z_n - d)}{2F^2}x^2)$$
$$\times \int_{-\infty}^{\infty} E_n(\xi)\exp(\frac{ikx\xi}{F})d\xi \qquad (6)$$

where $k = 2\pi/\lambda$. $\Delta z_n$ is the longitudinal position shift of the nth virtual source given by

$$\Delta z_n = n*2t\cos\theta_i \qquad\qquad (7)$$

Substituting eqn (4) in eqn (6) and making the transformation $\xi' = \xi + L$, eqn (6) can be written as

$$E_{out}(x,k) \propto \exp(ikd)\exp(\frac{-ikL}{F}x)\exp(\frac{-ikd}{2F^2}x^2)\sum_{n=0}^{\infty}\exp(-ik\Delta z_n)\exp(\frac{ik\Delta z_n}{2F^2}x^2)$$
$$\times \int_{-\infty}^{\infty} (R)^n E_0 \exp(-\frac{(\xi'-\Delta\xi_n)^2}{w_0^2})\exp(\frac{ikx\xi'}{F})d\xi' \qquad (8)$$

We would like to look at the above equation as being constituted of two relevant terms. The expression inside the summation corresponds to the demultiplexing property of the VIPA with the L=0, d=0 condition as discussed in [17]. However the introduction of the new parameters has introduced certain extra spatial phase terms, a linear phase term corresponding to L and as expected a quadratic phase term corresponding to d. Now in order to obtain the introduced spectral phases we will go through the following scheme. The demultiplexer part of the equation (included in the summation) gives us the wavelength to space mapping in the Fourier plane; using this mapping we will evaluate the spectral phases arising from the additional spatial phases.

From using the simplification for the demultiplexer part [17], we have the expression as –

$$E_{out}(x,k) \propto \exp(ikd)\exp(\frac{-ikL}{F}x)\exp(\frac{-ikd}{2F^2}x^2)$$

$$\exp(\frac{-x^2}{w'^2})\left(\frac{1}{1-\mathrm{R}\,exp(-ik(2t\cos\theta_i - 2t\sin\theta_i\frac{x}{F} - t\cos\theta_i\frac{x^2}{F^2}))}\right)\ (9)$$

Where $w'$ is an effective width of the intensity envelope on the Fourier plane. This depends on the input beam size and the focal lengths of other lenses used in the system. Fig 9 shows a schematic of the spectral dispersion by the VIPA. Each wavelength is dispersed into multiple orders with the relative powers between orders decided by the Gaussian term in eqn (9) which gives a Gaussian intensity envelope. More details on this can be obtained in [17].

From eqn (9) we have the additional spatial phase terms as

$$\exp(-ikL\frac{x}{F})\exp(-ikd\frac{x^2}{2F^2}) \qquad (10)$$

Here we have neglected the constant phase term since it does not contribute to any observable phenomenon.

From the VIPA demultiplexer part, we see that the equation for the VIPA demultiplexer, which gives us the positions for a given wavelength becomes-

$$k(2t\cos\theta_i - 2t\sin\theta_i\frac{x}{F} - t\cos\theta_i\frac{x^2}{F^2}) = 2m\pi \qquad (11)$$

i.e. $2t\cos\theta_i - 2t\sin\theta_i\frac{x}{F} - t\cos\theta_i\frac{x^2}{F^2} = m\lambda \qquad (12)$

$\lambda$ is the wavelength and $m$ is the order of diffraction. We clearly see the nonlinear relation connecting the position and wavelength. It is necessary here to comment on how the diffraction order $m$ is obtained. As shown in fig 9, we see that the nature of spectral dispersion by the VIPA causes the intensity profile on the Fourier plane to be a Gaussian centered around x=0. So, for a given frequency the diffraction order which disperses it nearest x=0 will have the most intensity. For a given wavelength $\lambda$, we have from eqn (12) that the diffraction order given by $m = \{2t\cos\theta_i / \lambda\}$ ({} representing closest integer) causes it to disperse around $x = 0$ (obtained by substituting x=0 in the equation) and hence has the highest intensity. Given a wavelength, we will refer to the order which has the highest intensity as the main order. Let us also denote the center wavelength of each order as $\lambda_m = 2t\cos\theta_i / m$ (which would fall on x=0). A center wavelength which is an FSR higher or lower (i.e. $\lambda_{m\pm1} = \lambda_m \pm \Delta\lambda$, where $\Delta\lambda$ is the FSR) will also correspond to $x = 0$; the main order in this case will be $m\pm1$, and so on.

Since a linear spatial phase term also leads to higher order spectral phases, an obvious point would be: why don't we just work at L=0 and set $d$ =0 to avoid any quadratic terms? To explain why we need L>0, we have pictorially represented the three possible cases in Fig. (10). For the power to couple back into the VIPA, it is necessary to have overlap between the VIPA and the reflected virtual sources. The optical system inverts the transverse positions of the sources; consequently overlap is only possible when L>0. Figures 10(A) and 10(B) shows the cases for L=0 and L< 0, for which we see no overlap. For L>0, shown in Fig. 10(C), we do get overlap. Depending on parameters such as the input angle and the VIPA design, there is an optimal L which gives maximum overlap. Though we discussed this using a reflective configuration, this argument is equally valid for a transmissive geometry (with a second lens and VIPA). In either VIPA pulse shaper configuration, the first virtual source on the output side cannot be at the same position as the first virtual source on the input side, which means a linear spatial phase is always present. However, the coefficients of the phase depend on the specific experimental setup.

Eqn (9) which gives the spatial phases is still incomplete. We need to incorporate two more factors. 1) A pulse shaper is a dual pass system, and by reciprocity the magnitudes of the phases need to be doubled. 2) In eqn (9) only the parameters $L$ and $d$ are present and we still haven't incorporated the mirror angle $\theta_m$. To do this let us look at the physical effect of a mirror with a small tilt. What the mirror does is tilt the wavefront by an angle twice the mirror angle causing an additional linear spatial phase given by

$$\exp(-ik\tan 2\theta_m x) \quad (13)$$

Hence $\theta_m$ has a similar effect as $L$. This can also be looked at as effectively causing a transverse offset of the reflected virtual sources by an amount $F \tan 2\theta_m$. In this case we don't need to multiply by two since it is only in the return path. A way of visualizing this is, for L>0 and $\theta_m = 0$, the distance between the incident and reflected first virtual source is 2L, but for L=0, $\theta_m \neq 0$ it will only be $F \tan 2\theta_m$ and not twice that. A rigorous derivation of this can be obtained by looking at the expressions for the wavefronts but the essential idea is as we described above. Including these effects, the expression for the spatial phase becomes

$$\exp(-ik(2L + F \tan 2\theta_m)\frac{x}{F})\exp(-ikd\frac{x^2}{F^2}) \qquad (14)$$

To obtain the expressions for spectral phases, first let us solve for x from eqn (12) in terms of $\lambda$ and then substitute the result into eqn (14). The solution for eqn (12) is

$$\frac{x}{F} = \frac{-\sin\theta_i + \sin\theta_i \sqrt{1 - \dfrac{m(\lambda - \lambda_m)\cos\theta_i}{t\sin^2\theta_i}}}{\cos\theta_i} \qquad (15)$$

Where $\lambda_m = 2t\cos\theta_i / m$ is the center wavelength for order m and we have chosen the solution closer to x=0, which as we discussed before is the relevant solution due to the intensity envelope. For a given order m, the maximum excursion for the term $(\lambda - \lambda_m)$ is $\Delta\lambda / 2$ where $\Delta\lambda$ is the FSR. So we have,

$$\frac{m(\lambda - \lambda_m)\cos\theta_i}{t\sin^2\theta_i} < \frac{m\Delta\lambda\cos\theta_i}{2t\sin^2\theta_i} \sim \frac{\lambda\cos\theta_i}{2t\sin^2\theta_i} \qquad (16)$$

Where we have used $m \sim 2t\cos\theta_i/\lambda$. For the space of operating parameters in our experiment, which is $\theta_i \sim 4$degrees and t=0.75mm, we have

$$\frac{m(\lambda - \lambda_m)\cos\theta_i}{t\sin^2\theta_i} < \frac{\lambda\cos\theta_i}{2t\sin^2\theta_i} \sim 0.2 \qquad (17)$$

Since this is reasonably smaller than one in our case and in general for a significant range of operating parameters, we will expand the expression inside the square root sign using the binomial expansion rule $\sqrt{1-\alpha} = 1 - \alpha/2 - \alpha^2/8 - \alpha^3/16 - 5\alpha^4/128....$ [18] up to the 4$^{\text{th}}$ order term. This gives us

$$\frac{x}{F} \sim -\frac{1}{2t\sin\theta_i}m(\lambda - \lambda_m) - \frac{\cos\theta_i}{8t^2\sin^3\theta_i}m^2(\lambda - \lambda_m)^2 - \frac{\cos^2\theta_i}{16t^3\sin^5\theta_i}m^3(\lambda - \lambda_m)^3 - \frac{5\cos^3\theta_i}{128t^4\sin^7\theta_i}m^4(\lambda - \lambda_m)^4$$

$$(18)$$

Using this expression in the expression for the spatial phases (eqn (14)), we group the spectral phases according to different powers of $(\lambda - \lambda_m)$ as follows:

Linear phase -

$$\exp(ik\frac{(L + \frac{F}{2}\tan 2\theta_m)}{t\sin\theta_i}m(\lambda - \lambda_m)) \qquad (19)$$

Quadratic phase -

$$\exp(ik\frac{(L+\frac{F}{2}\tan 2\theta_m - d\tan\theta_i)\cos\theta_i}{4t^2\sin^3\theta_i}m^2(\lambda-\lambda_m)^2) \qquad (20)$$

Cubic phase -

$$\exp(ik\frac{(L+\frac{F}{2}\tan 2\theta_m - d\tan\theta_i)\cos^2\theta_i}{8t^3\sin^5\theta_i}m^3(\lambda-\lambda_m)^3) \qquad (21)$$

Biquadratic phase –

$$\exp(ik\frac{(L+\frac{F}{2}\tan 2\theta_m - d\tan\theta_i)5\cos^3\theta_i}{64t^4\sin^7\theta_i}m^4(\lambda-\lambda_m)^4) \qquad (22)$$

A point to note here is that the presence of '$k = 2\pi/\lambda$' in the above expression also contributes higher order terms owing to the '$1/\lambda$' dependency. However it can be easily shown that the higher order terms due to this are significantly smaller than the higher order terms in the spectral phase given above; hence we will neglect this contribution and use the expressions given above.

We are not interested in the linear spectral phase since that is just a pulse delay in the time domain, but 2$^{nd}$ and higher order spectral phases affect the temporal envelope of the pulse. Along with 2$^{nd}$ order spectral phase, we also have higher order spectral phases which cause pulse distortion. We notice that all of eqns (20-22), which give the expressions for the quadratic, cubic and biquadratic phases, contain the common coefficient $L+\frac{F}{2}\tan 2\theta_m - d\tan\theta_i$. In fact, we will now show that this coefficient is in every term of degree>1. From eqn (15) we have $\frac{x}{F}$ in the functional form –

$$\frac{x}{F} = \tan\theta_i(\sqrt{1-\alpha} - 1) \qquad (23)$$

where $\alpha = \dfrac{m(\lambda - \lambda_m)\cos\theta_i}{t\sin^2\theta_i}$. Substituting this in the equation for the spatial phases (eqn (14)), we

have the expression for spectral phases as

$$e^{-i2k\tan\theta_i(L+\frac{F}{2}\tan 2\theta_m - d\tan\theta_i)(\sqrt{1-\alpha}-1)} \; e^{ikd\tan^2\theta_i\alpha} \qquad (24)$$

It is clear from the above the expression that when the expression is expanded in a power series in terms of $\alpha$, which is a constant multiple of $(\lambda - \lambda_m)$, for all orders of $\alpha > 1$, the coefficient will contain the term $L + \dfrac{F}{2}\tan 2\theta_m - d\tan\theta_i$. If we can adjust the parameters of the experiment to

make this term zero, then all 2$^{\text{nd}}$ and higher order phases will go to zero. This will provide us a

configuration with no dispersion or pulse distortion and will be the true zero dispersion condition

for a VIPA based pulse shaper. So we have proved the zero dispersion condition we discussed in

section 3. Stating again we have -

Zero dispersion condition

$$L + \frac{F}{2}\tan 2\theta_m - d\tan\theta_i = 0 \qquad (3)$$

It is interesting to look at the physical interpretation of the zero dispersion condition, now in the spectral phase perspective. By nature of the VIPA based pulse shaper configuration, a linear spatial phase is unavoidable and hence the nonlinear sampling of this linear phase will force higher order spectral phase terms. However the distance between the VIPA and Fourier lens which introduces a quadratic spatial phase, upon nonlinear sampling, will also introduce second and higher order spectral phases. So what we are doing here is to change the VIPA-lens distance to create phases which are of opposite sign of those arising due to the spatial linear phase. When the phases arising from these two different mechanisms are equal and opposite, net zero dispersion configuration is achieved.

The constants in equation (3) give us an idea of the importance of individual parameters. Since the input angle to the VIPA is usually small (< 5deg), and the focal length of the Fourier lens is tens of centimeters, we see that the temporal dispersion is much more sensitive to $L$ and $\theta_m$ than it is to $d$. As an example, for a VIPA with a 200GHz FSR, an input angle of 4 degrees and a reflectivity of 98% with $\theta_m = 0$, the value of L to obtain minimum loss is ~ 2.6mm [19]. To compensate the temporal dispersion that results, the distance the lens has to be moved from the 4-F configuration is ~3.7 cm. In our experiments this is ~10% of the 40 cm focal length of the lens.

## 5. Experimental verification and spectral monitoring of temporal dispersion

In the previous section we modeled the dispersion behavior in VIPA based pulse shapers and provided expressions for the spectral phases. In this section we will describe some experimental results that verify our model. In particular, we measure the temporal dispersion introduced as relevant experimental parameters are varied and compare the results with the theoretical predictions. Dispersion can be measured in various ways either directly using pulse measurement techniques like FROG [20] or indirectly by looking at intensity cross-correlations of the output with a short reference pulse. The latter is a common methodology used for pulse shapers owing to comparative simplicity of the apparatus and the known nature of the input pulse. However, here we will adopt a new approach in which we obtain the temporal dispersion through the spectral interference between different diffraction orders. Utilizing this approach has two key benefits – it is very easy to implement, and it provides a real time diagnostic of the dispersion properties which is convenient for tuning the pulse shaper to the zero dispersion configuration.

Let us first quantitatively describe the spectral interference effect. Let $\psi^m(\lambda)$ and $\psi^{m+1}(\lambda)$ be the spectral phase for a wavelength $\lambda$ for two adjacent diffraction orders 'm' and 'm+1'. Since the intensity distribution at the Fourier plane follows a Gaussian envelope (fig 9), interference effects will largely be limited to two adjacent diffraction orders dispersed around x=0. This justifies in modeling with only two orders. The output power at wavelength $\lambda$ now depends on the coherent sum of contributions from two orders, which depends on their phase difference. If the phase difference is a function of $\lambda$, then as the wavelength changes the interference condition may change from constructive to destructive, causing spectral amplitude ripples. First taking into consideration only the linear spectral phase (eqn 19) we have

$$\psi^{m+1}{}_{linear}(\lambda) - \psi^m{}_{linear}(\lambda) = \frac{2\pi}{\lambda} \frac{(L + \frac{F}{2}\tan 2\theta_m)}{2t\sin\theta_i}[(m+1)\lambda - 2t\cos\theta_i - m\lambda + 2t\cos\theta_i]$$

(25)

$$\Rightarrow \psi^{m+1}{}_{linear}(\lambda) - \psi^m{}_{linear}(\lambda) = 2\pi \frac{(L + \frac{F}{2}\tan 2\theta_m)}{2t\sin\theta_i}$$

(26)

In the simplification we have used $(m+1)\lambda_{m+1} = m\lambda_m = 2t\cos\theta_i$ based on the definition of 'm' as discussed before. We see that the phase difference is independent of $\lambda$ and hence does not contribute to spectral ripples. However if we take the quadratic spectral phase (eqn 20) we have

$$\psi^{m+1}{}_{quadratic}(\lambda) - \psi^m{}_{quadratic}(\lambda) =$$

$$\frac{2\pi}{\lambda}\frac{(L+\frac{F}{2}\tan 2\theta_m - d\tan\theta_i)\cos\theta_i}{4t^2\sin^3\theta_i}[((m+1)\lambda - 2t\cos\theta_i)^2 - (m\lambda - 2t\cos\theta_i)^2] \qquad (27)$$

$$\Rightarrow \psi^{m+1}{}_{quadratic}(\lambda) - \psi^m{}_{quadratic}(\lambda) =$$

$$2\pi\frac{(L+\frac{F}{2}\tan 2\theta_m - d\tan\theta_i)\cos\theta_i}{4t^2\sin^3\theta_i}[(2m+1)\lambda - 4t\cos\theta_i] \qquad (28)$$

where we have again used the fact that $(m+1)\lambda_{m+1} = m\lambda_m = 2t\cos\theta_i$. We see that the phase difference between adjacent orders is linearly related to $\lambda$, so as $\lambda$ changes the coherent sum is expected to change. Since for the case of quadratic spectral phase, the phase difference is linearly proportional to $\lambda$, the leading interference term between adjacent orders corresponds to a sinusoidal spectral ripple with a constant period. If we define the period as P, then

$$[\psi^{m+1}(\lambda+P) - \psi^m(\lambda+P)] - [\psi^{m+1}(\lambda) - \psi^m(\lambda)] = 2\pi \qquad (29)$$

Substituting this in eqn (28) and again making use of $m = 2t\cos\theta_i / \lambda_m$, we have

$$2\pi\frac{(L+\frac{F}{2}\tan 2\theta_m - d\tan\theta_i)\cos\theta_i}{4t^2\sin^3\theta_i}[(2m+1)P] = 2\pi \qquad (30)$$

With the approximation $2m \gg 1$, the expression for the constant ripple period is

$$P \approx \frac{\lambda_m t \sin^3 \theta_i}{(L + \frac{F}{2} \tan 2\theta_m - d \tan \theta_i) \cos^2 \theta_i} \qquad (31)$$

Similarly looking at higher order spectral phases will show phase differences which vary with wavelength at power 2 and higher. However, the magnitudes of phase contributions progressively reduce as we go to higher powers of wavelength. Hence we can visualize the spectral ripple formation as a constant period function due to the 2nd order phase with the period modified to some extent by the higher order spectral phases (see for example fig 2(D)). In a regime where we are close to zero dispersion, the 2nd order term will dominate and we expect reasonably constant period ripples. We will use this regime for the experimental verification of our model. Although we are not directly verifying the coefficients of 3rd and higher order spectral phases, we are directly verifying the most important equation - the zero dispersion condition essential for the operation of the pulse shaper. For configurations far away from zero dispersion where higher order phases are important, the fact remains that fewer spectral ripples imply lower dispersion and can be used as a guide to approach the zero dispersion setting.

In the zero dispersion condition (eqn 3), we have three different parameters, namely $L$, $\theta_m$ and $d$, which if unbalanced contribute to temporal dispersion. Since measuring each individual parameter in absolute terms is difficult, for the experiment we adopted the following scheme. First, a zero dispersion condition was achieved by adjusting different parameters simultaneously with guidance from eqn (3). Zero dispersion was verified both by achieving a flat featureless spectrum and by measuring the intensity cross-correlation to show that the output was a short bandwidth limited pulse. Then each of the parameters $L$, $\theta_m$ and $d$ were individually varied by known quantities while the other two were fixed, and the average spectral ripple period as obtained through optical spectrum analyzer measurements was recorded. The experimental results were compared to the prediction of our theory (eqn (31)).

For a change in transverse length of $\Delta L$ from the zero dispersion condition, from eqn (31) we have the period of ripple as –

$$P \sim \frac{\lambda_m t \sin^3 \theta_i}{\cos^2 \theta_i} \frac{1}{\Delta L} \tag{32}$$

Similarly, a change in the mirror angle by $\Delta \theta_m$ or a change in the VIPA – lens distance by $\Delta d$ from the zero dispersion condition gives the expressions for the ripple period as

$$P \sim \frac{\lambda_m t \sin^3 \theta_i}{\cos^2 \theta_i} \frac{1}{\frac{F}{2}\Delta(\tan 2\theta_m)} \sim \frac{\lambda_m t \sin^3 \theta_i}{\cos^2 \theta_i} \frac{1}{F \Delta \theta_m} \tag{33}$$

$$P \sim \frac{\lambda_m t \sin^2 \theta_i}{\cos \theta_i} \frac{1}{\Delta d} \tag{34}$$

In eqn 33, we have made use of the fact that the mirror tilt in our experiments is so small (<0.5deg) that we can approximate $\tan 2\theta_m = 2\theta_m$. Fig 11 shows the calculated and the experimentally observed ripple periods as each parameter is varied. We see an excellent agreement between the two. In these experiments an air filled VIPA with a design wavelength of 1550nm and a FSR of 200GHz was used. The input angle to the VIPA was 4 degrees and the focal length of the Fourier lens used was 40 cm. In these plots we have adjusted the alignment parameters over ranges that cause the observed ripple periods to vary over approximately the same range. We see that a change in d of 2.2 cm will have the roughly the same effect as changing L by 1.5mm or changing the mirror angle by 0.21 degrees. This highlights the importance of precisely tuning the L and $\Delta\theta_m$ parameters.

In pulse shapers, as mentioned previously, the amount of temporal dispersion present is usually determined by directly measuring the spectral phase or by taking cross-correlations with a bandwidth limited reference signal. To achieve zero dispersion, this process may be performed repetitively while changing the alignment parameters. However, this is a time consuming procedure, especially for VIPAs which have increased alignment degrees of freedom compared to gratings. Alternately, when a spatial light modulator is placed at the Fourier plane, the spatial light modulator may be programmed to compensate the dispersion. However, for large dispersion this wastes the phase dynamic range available. In contrast, the spectral ripples discussed above make possible a very convenient method to monitor the temporal dispersion directly in the frequency domain. By monitoring the period of output spectral ripples on an optical spectrum analyzer working in sweep mode with sufficient resolution (sufficiently smaller than the FSR of the VIPA), the temporal dispersion can be quickly calculated. Any time-dependent variations in dispersion would give rise to changes in the ripple period, allowing measurement with fast update limited only by the sweep time of the OSA. Also, obtaining zero dispersion becomes more convenient since the monitoring can be done purely in the spectral domain by looking for a flat output spectrum.

Fig 12(A) – (E) show the variation of a one FSR section of the spectrum as the transverse position (L) of the lens with respect to the VIPA is varied moving towards zero dispersion. These experiments were performed using a 2D VIPA-grating hybrid pulse shaper [13], which has a behavior similar to a VIPA only shaper at these bandwidths. The input angle to the VIPA was ~4 degrees. We see that the ripple period continuously increases (ripple frequency continuously decreases) and at zero temporal dispersion the spectrum is flat (fig 12(E)). At this configuration the cross-correlation was also obtained (fig 12(F)), and a bandwidth limited output pulse is observed indicating the absence of temporal dispersion. Here we would like to contrast the results of Figs. 12(E,F) with that seen in Figs. 2(C,D) where in the presence of dispersion a long pulse train was observed in the time domain, and strong amplitude ripples were observed in the spectral domain.   The relationship between temporal dispersion and spectral amplitude ripples is not limited to VIPAs but applies to any spectral disperser which shows diffraction into multiple orders and has a significant nonlinear frequency to space mapping.

A point we would like to stress is that temporal dispersion is a spectral phase effect which in presence of multiple orders creates spectral amplitude effects. In many situations it may be convenient to use only one order and to block all the others. In this case the spectral amplitude manifestation of dispersion will disappear but temporal broadening due to the dispersion will remain.

Before we end this section, we comment on the shape of the spectral ripples. Two main factors affect this - the relative intensities of the two interfering orders and the contributions from higher order phases. An exact solution is difficult to obtain and also is not very useful. However it is useful to understand the effects qualitatively.   We have already discussed the effect of higher order phases as modifying the ripple periods. The wavelength dependent distribution of intensity between orders does not affect the period of the ripples but affects the fringe contrast.   At wavelengths for which most of the power is in a single order (at the center of the FSR), the ripple contrast is very low, while at the edges of each FSR where two adjacent orders have equal powers ,the ripple contrast is maximum. This effect can be clearly seen in Fig. 2(D).

## 6. Relation between loss and dispersion for VIPA pulse shapers

As we discussed previously with the help of fig 10, the loss of VIPA based pulse shapers depends on the overlap between the propagating beam and the returning beam. For a given VIPA, this overlap can be represented as a function of the distance between the first virtual source in the propagating beam and the position of the first virtual source in the returning beam [19]. The optimal value for this parameter will depend on other parameters like the physical dimensions of the VIPA and the reflection coefficient per bounce inside the VIPA, but for a given setup when all these other parameters are fixed, the distance parameter is sufficient to evaluate the overlap.

For the general reflective VIPA shaper configuration as shown in Fig. 10, the distance between the first virtual source in the forward beam and in the returning beam can be written as $2(L + \frac{F}{2}\tan 2\theta_m)$. This quantity has to be greater than zero to assure some overlap and ideally should be the optimal value for the VIPA being used. When we compare this with eqn (3) which characterizes the zero dispersion condition, we see that the first two terms are the same. This indicates the close connection between the loss and the zero dispersion condition for VIPA pulse shapers. For any d > 0, the values for $L$ and $\theta_m$ can be chosen appropriately to obtain zero dispersion, but this may take us away from the minimum loss condition. So an algorithmic approach is necessary to achieve low loss and zero dispersion for the shaper at the same time. First, suitable values for the parameters $L$ and $\theta_m$ need to be selected to achieve maximum overlap and hence minimum loss. Then using these values the value of d should be selected to achieve zero dispersion. This will ensure that the low loss and zero dispersion conditions can be achieved simultaneously.

Another point to notice is that when $d \leq 0$ (i.e. the VIPA – lens distance is greater than the focal length), to achieve zero dispersion we would require $L + \dfrac{F}{2} \tan 2\theta_m < 0$ which would lead to zero overlap and hence infinite loss. This implies that a VIPA shaper does not have zero dispersion configuration for $d \leq 0$, and as a corollary the 4-F configuration for a VIPA shaper cannot be dispersion free.

## 7. Summary and conclusions

In summary, we have demonstrated new aspects of temporal dispersion applicable to pulse shapers which use spectral dispersers with a nonlinear frequency to space mapping. For the case of VIPA based pulse shapers, we provided a rigorous quantitative description followed by experimental verification. Our results reveal that such pulse shapers exhibit strong temporal dispersion in the 4F configuration, which is normally considered to be dispersion free. Zero dispersion may still be obtained, but only with substantial change of the spectral disperser – lens separation. We also showed a close connection between the loss and the dispersion of VIPA pulse shapers; understanding this connection facilitates achieving a pulse shaper configuration which simultaneously has low loss and zero dispersion. Finally, we demonstrated that temporal dispersion – coupled with contributions from multiple diffraction orders – can lead to strong interference fringes in the power spectrum, an effect which can be used for convenient monitoring of the dispersion of the setup.

These effects should not be limited to VIPA based pulse shapers. For example, we expect some of these effects to become significant for grating based configurations when the optical bandwidth becomes large (e.g., few cycle pulses), such that the nonlinear spectral dispersion of the grating becomes more pronounced and multiple diffraction orders from the grating begin to fall within the spectrum.

## Acknowledgements

We would like to thank the NSF under grant ECCS-0601692 and DARPA/AFOSR under grant FA9550-06-1-0189. We would also like to thank Avanex Corporation for the VIPA devices.

Fig. 1 Schematic showing spectral dispersion by a VIPA due to interference between multiple virtual sources.

Fig. 2 (A) input pulse, (B) input spectrum, (C) output pulse, (D) output spectrum. Note that spectra are plotted only over one free spectral range of the VIPA.

Fig. 3 (A) schematic showing the different parameters varied (the lens to VIPA distance by a few cm, the transverse position of the VIPA by a few mm and the mirror tilt angle by a few tenths of a degree). (B, D, F) and (C, E, G) are respectively the time domain cross-correlations and the corresponding spectra as the transverse position of the lens is varied. Similar behavior is observed with the variation of any of the three parameters. Spectra are plotted only over one free spectral range of the VIPA.

Fig 4 Figure showing the origin of temporal dispersion in grating and VIPA pulse shapers. (A) Temporal dispersion in a grating pulse shaper when not in 4-F configuration. (B) Nature of spectral dispersion by the VIPA due to interference between multiple virtual sources. (C) Location of the incident and reflected virtual sources in the 4-F configuration indicating the origin of additional dispersion. (D) Configuration showing the overlap between incident and reflected virtual sources – expected zero dispersion condition for VIPA pulse shapers.

Fig. 5 Spectral dispersion of a VIPA (from [17])

Fig. 6 Schematic showing creation of higher order spectral phases due to nonlinear sampling of spatial linear phase, where different spots represent equally spaced frequencies. (A) For linear

spectral dispersers, the phase differences between equally spaced frequencies are constant. (B) For nonlinear spectral dispersers, the phase differences between equally spaced frequencies are now frequency dependent (leading to quadratic and higher order spectral phase).

Fig 7 A spatial linear phase sampled nonlinearly in presence of multiple diffraction orders.

Fig. 8 A general configuration of a VIPA pulse shaper in the reflection geometry.

Fig. 9 A schematic representation of spectral dispersion by the VIPA. Light is spectral dispersed into multiple orders under a Gaussian intensity envelope.

Fig. 10 Figure showing overlap conditions between input and reflected sources for varying L. (A) L=0, no overlap. (B) L<0, no overlap. (C) L>0, overlap.

Fig. 11 Calculated and experimentally observed ripple periods for (A) variation in L, (B) variation in $\theta_m$, (C) variation in d.

Fig. 12 (A) – (E) Observed spectrum of one VIPA free spectral range as the transverse position (L) of the lens is adjusted (while VIPA lens distance (d) and mirror angle ($\theta_m$) are maintained fixed). L is adjusted to move monotonically towards zero dispersion. (F) Cross-correlation signal showing time domain output of the VIPA pulse shaper at zero dispersion.

Fig. 1

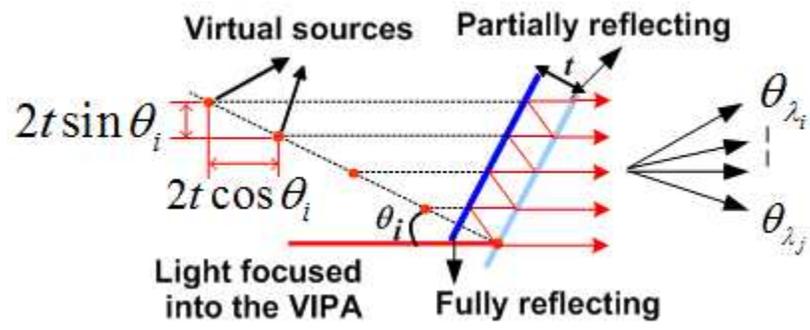

Fig. 2

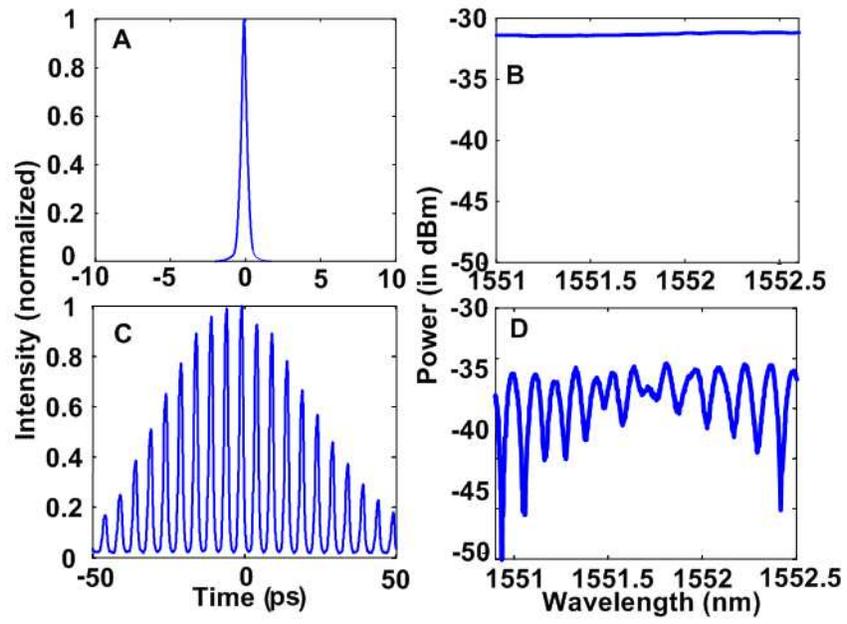

Fig. 3

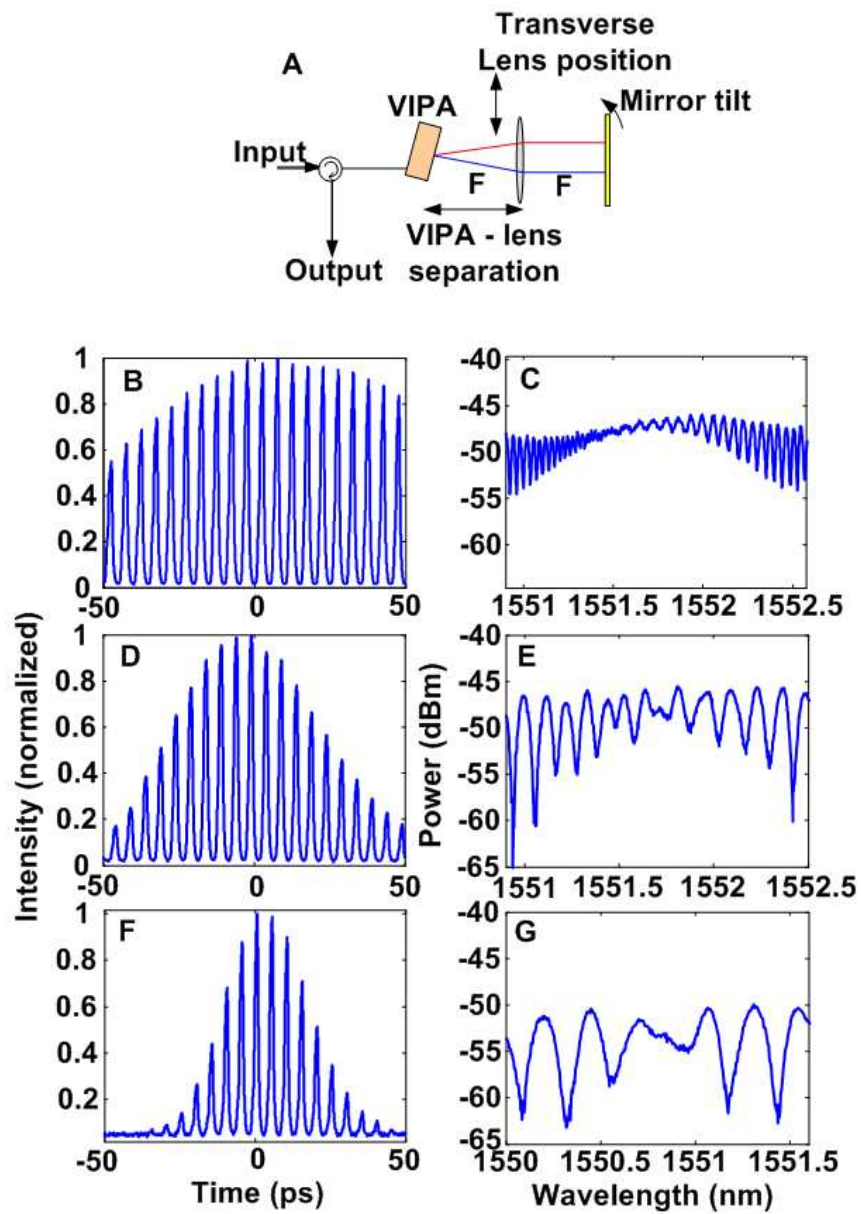

Fig. 4

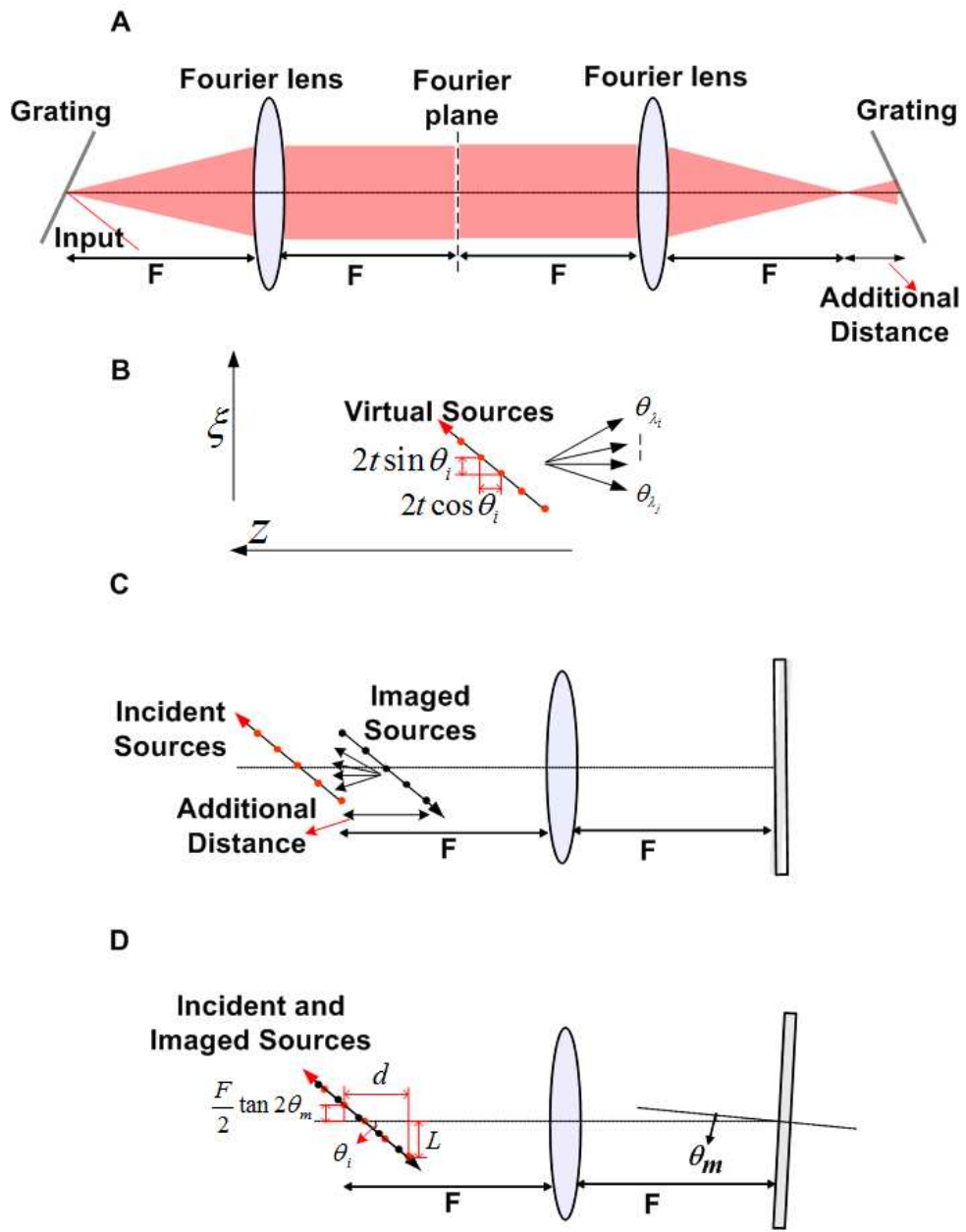

Fig. 5

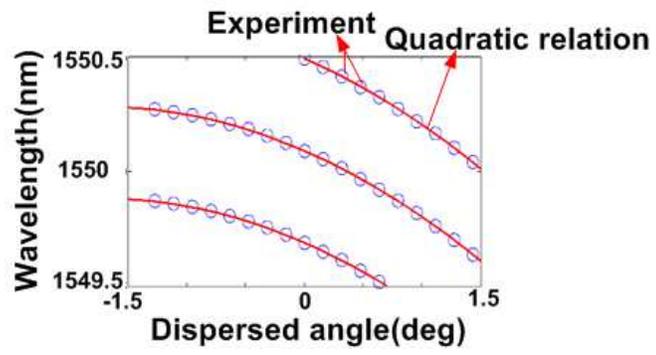

Fig. 6

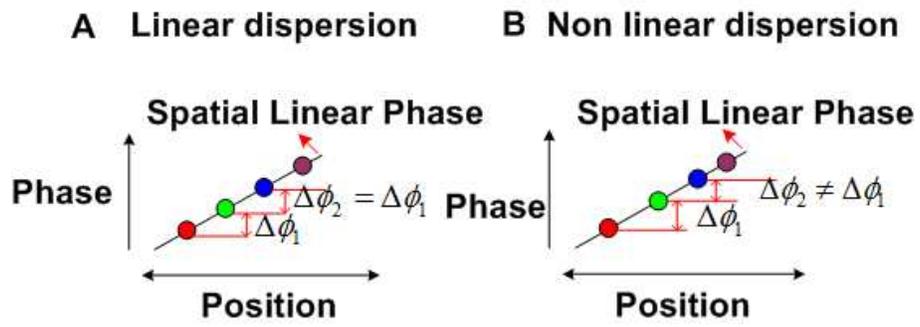

Fig. 7

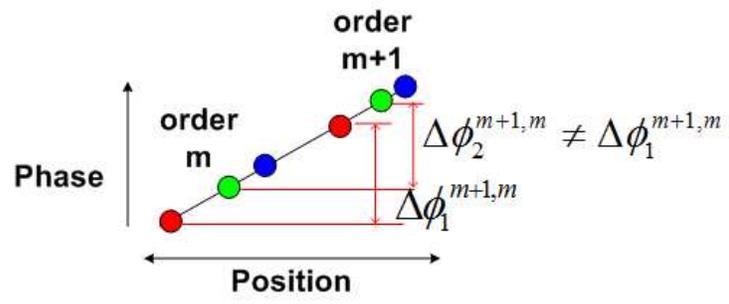

Fig. 8

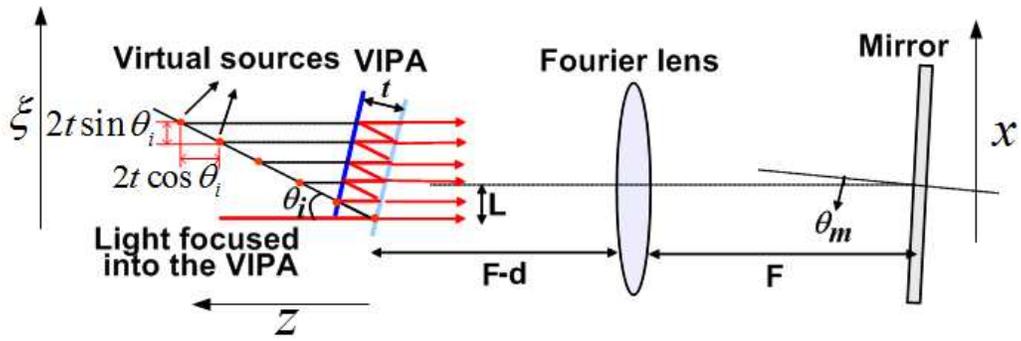

Fig. 9

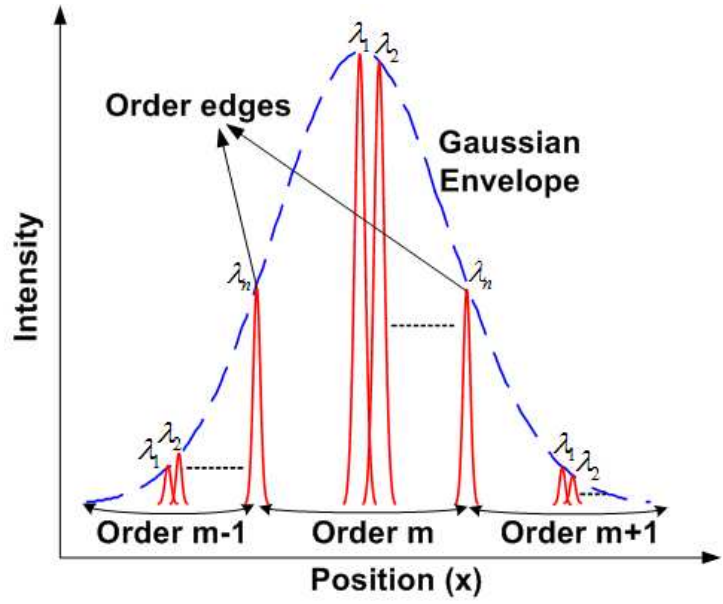

Fig. 10

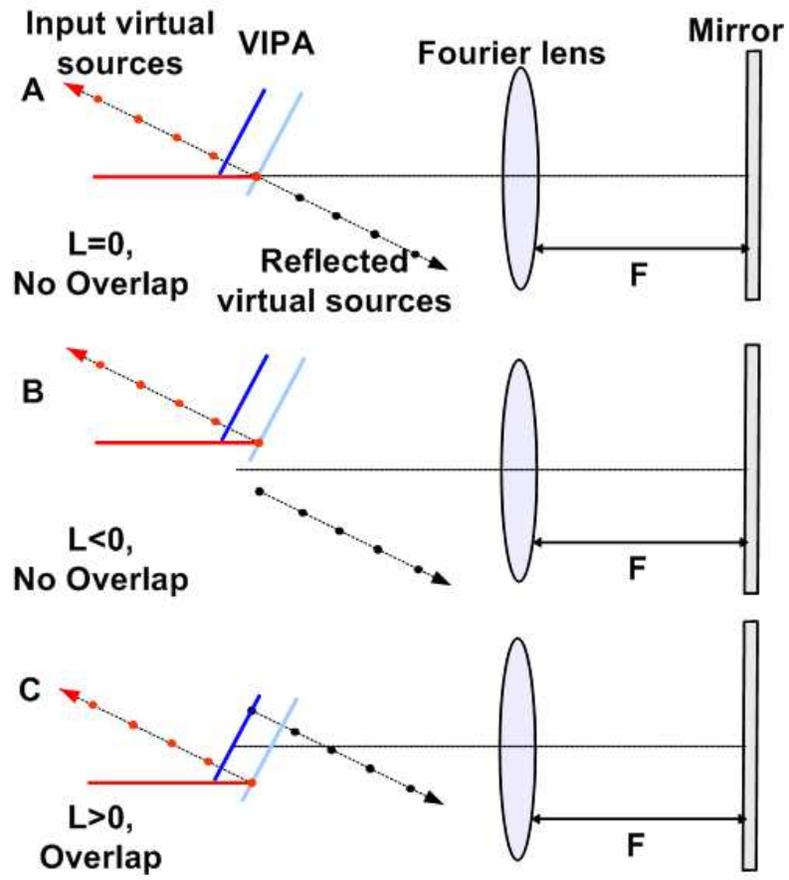

Fig. 11

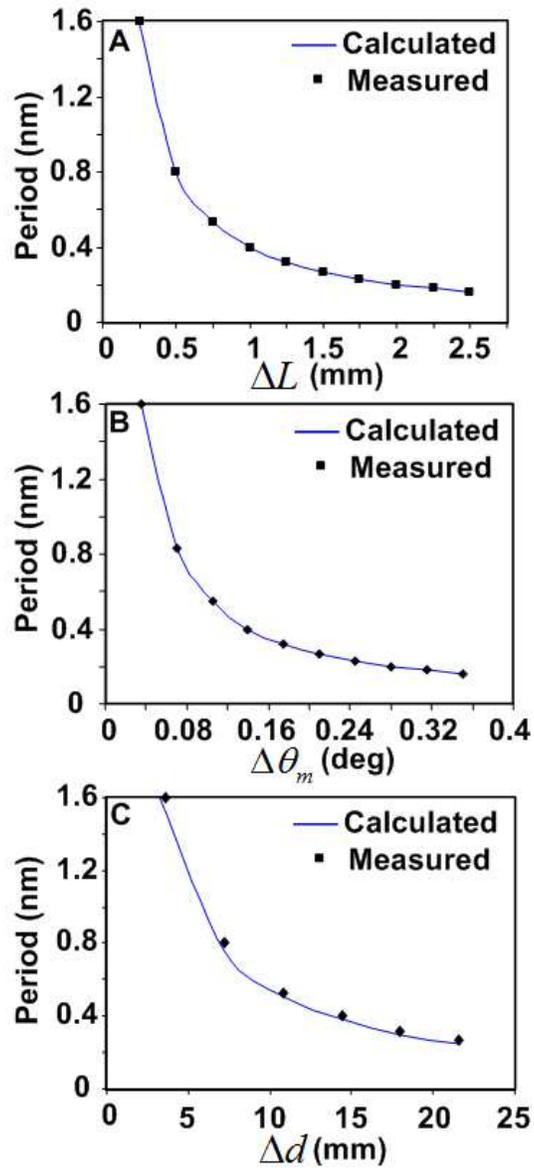

Fig. 12

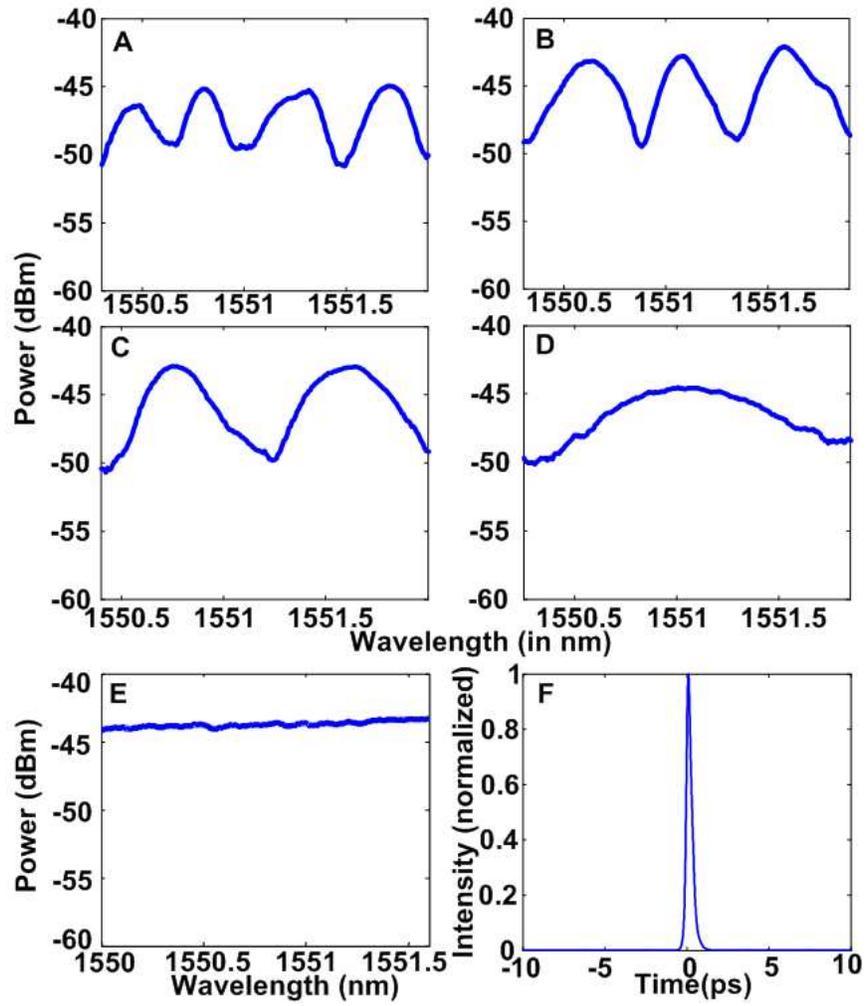